\documentclass[
    ,final            
  ]
  {aipproc}

\layoutstyle{8x11single}

\begin{document}

\title{Electromagnetic Polarizabilities: \\ Lattice QCD in Background Fields}


\author{W.~Detmold}{
  address={College of William and Mary, Williamsburg, VA}
  ,altaddress={Thomas Jefferson National Accelerator Facility, Newport News, VA} 
}

\author{B.~C.~Tiburzi}{
  address={Center for Theoretical Physics, Massachusetts Institute of Technology, Cambridge, MA}
}

\author{A.~Walker-Loud}{
  address={Lawrence-Berkeley National Laboratory, Berkeley, CA}
}

\begin{abstract}
Chiral perturbation theory makes definitive predictions for the extrinsic behavior of hadrons in external electric and magnetic fields. 
Near the chiral limit, 
the electric and magnetic polarizabilities of pions, kaons, and nucleons are determined in terms of a few well-known parameters. 
In this limit, 
hadrons become quantum mechanically diffuse as polarizabilities scale with the inverse square-root of the quark mass.
In some cases, however, 
such predictions from chiral perturbation theory have not compared well with experimental data. 
Ultimately we must turn to first principles numerical simulations of QCD to determine properties of hadrons,
and confront the predictions of chiral perturbation theory. 
To address the electromagnetic polarizabilities, 
we utilize the background field technique. 
Restricting our attention to calculations in background electric fields, 
we demonstrate new techniques to determine 
electric polarizabilities and baryon magnetic moments for both charged and neutral states. 
As we can study the quark mass dependence of observables with lattice QCD, 
the lattice will provide a crucial test of our understanding of low-energy QCD, 
which will be timely in light of ongoing experiments, such as at COMPASS and HI$\gamma$S. 
\end{abstract}

\maketitle


\section{Motivation}

Beyond intrinsic properties of hadrons, which characterize their internal structure, 
there are extrinsic properties which characterize the response of a hadron to external conditions. 
Electric and magnetic polarizabilities are examples of such properties, 
and are the only measured extrinsic properties of hadrons listed in the PDG. 
The electric polarizability
$\alpha_E$, 
for example,  
characterizes the strength of the induced electric dipole moment, 
$\vec{p}_E$,
when the hadron is subjected to an external field, 
$\vec{p}_E = - \alpha_E \vec{E}$. 
For a hadron, 
$h$,
dimensional analysis gives us
\begin{equation}
\alpha_E(h)
= 
N(h)
\,
\alpha_{fs} \left( \frac{4}{3} \pi \, [ \texttt{fm}^3] \right)
,\end{equation}
where 
$N(h)$ is a pure number that is hadron dependent, 
and
$\alpha_{fs}$
is the fine-structure constant.

The MIT bag model~\cite{Chodos:1974je}, 
for example, 
provides a way to compute nucleon polarizabilities. 
The electric and magnetic polarizabilities one computes are the right order of magnitude~\cite{Schafer:1984hw}. 
Presumably this is due to the mechanism of confinement in the model which has as input a natural-sized hadronic length scale. 
For this model calculation, 
the quarks are taken to be massless. 
This \emph{chiral limit}, 
however,
is one in which the behavior of QCD can be understood using an effective theory~\cite{Gasser:1983yg}. 
Chiral perturbation theory results from considering the pattern of spontaneous and explicit symmetry breaking in low-energy QCD. 
This theory makes simple predictions for pion, kaon, and nucleon polarizabilities, 
which are generically of the form~\cite{Holstein:1990qy,Bernard:1991rq}
\begin{equation}
\alpha_E^\chi(h)
=
N^\chi(h)
\frac{\alpha_{fs}}{f_\pi  [m_q \langle \overline{\psi} \psi \rangle]^{1/2} }
,\end{equation}
where 
$f_\pi$ is the pion decay constant, 
$\langle \overline{\psi} \psi \rangle$
is the chiral condensate, 
and
$m_q$ is the quark mass. 
The pure numbers 
$N^\chi(h)$
are determined within chiral perturbation theory. 
The theory is only effective when there is a power counting to order the infinite tower of operators contained in the chiral Lagrangian. 
In general, 
the expansion is controlled by the smallness of the light quark masses compared to the chiral symmetry breaking scale. 
Despite numerous successes, 
the pion polarizability prediction is a factor of two different than the most recent experimental determination~\cite{Ahrens:2004mg}.

\section{Background Field Method}

To determine polarizabilities and confront the discrepancies with experiment, 
we turn to lattice QCD. 
Unfortunately direct evaluation of the Compton scattering tensor, 
$T_{\mu \nu} (\omega, \omega^\prime)$,  
with current state-of-the-art lattice computations is out of reach. 
One particular challenge that would need to be addressed is the quantization of momentum in units of 
$2 \pi / L$, 
due to the periodicity of the lattice. 
As such, 
one requires very large lattices,
$\omega, \omega^\prime = 2 \pi / L \ll m_\pi$,
to access the zero-momentum limit of the Compton tensor. 
A promising alternative is to use background fields~\cite{Bernard:1982yu,Martinelli:1982cb,Fiebig:1988en}.

The basics of the background field method are basic:
(i) study QCD in the presence of external fields by measuring correlation functions, 
and 
(ii) study the behavior of correlations functions to determine response parameters.  
To achieve the former, 
quarks are coupled to external electromagnetic fields through their charges. 
In practice, 
this amounts to multiplication of the color gauge links by a classical $U(1)$ link.\footnote{
This must be done for both valence quarks (affecting propagators) and sea quarks (affecting gauge field configurations).
Due to current computational restrictions, 
the sea quarks in our simulations are electrically neutral. 
We are investigating techniques to cure this malady. 
}
Only in the continuum limit are the quarks minimally coupled to the electromagnetic field. 
With gauge links, 
the uniform field strengths allowed on a lattice are subject to certain quantization conditions~\cite{'tHooft:1979uj}, 
else there is a considerably sized field gradient. 
Such huge gradients lead to undesirable effects, 
even when located far from the lattice measurements~\cite{Detmold:2008xk}.

To perform measurements in lattice QCD, 
one chooses an interpolating field
$\chi_h(x)$ 
having the quantum numbers of a desired hadron, $h$
(here we presume the ground state hadron is of interest). 
First, let us start with a neutral spin-less hadron operator. 
One computes the two-point correlation function in the applied field, 
$\mathcal{E}$, 
\begin{equation}
G_\mathcal{E} (\tau)
= 
\sum_{\vec{x}}
\langle 
\chi_h(\vec{x}, \tau) \chi^{\dagger}_h(0) 
\rangle_{\mathcal{E}}
= 
\sum_n 
Z_n(\mathcal{E}) 
e^{ - E_n(\mathcal{E}) \tau}
.\end{equation}
The long Euclidean time limit of the correlation function can be used to determine 
the ground-state hadron's energy 
$E_0 (\mathcal{E}) = M_h + \frac{1}{2} \alpha^h_E \mathcal{E}^2$, 
where we have kept only terms at second-order in the external field, 
which is presumed perturbatively small, 
and the sign of the second-order term arises from treatment in Euclidean space.   
Measurement of the two-point correlation function for several values of the external 
electric field, 
will allow one to determine the electric polarizability. 
We carried out such studies for the neutral pion (connected part) and neutral kaon~\cite{Detmold:2009dx}.

When one considers charged particles, 
this simple spectroscopic method will no longer work. 
We suggested that one could still extract useful information from the two-point correlation functions
of charged particles by matching onto the behavior predicted from single-particle effective actions~\cite{Detmold:2006vu}.  
For a charged spin-less particle subjected to the field 
$A_\mu = ( 0, 0, - \mathcal{E} x_4, 0)$, 
we have for $\vec{p} = 0$,
\begin{equation}
G^{-1} 
= 
- \frac{\partial^2}{\partial x_4^2}
+ 
\mathcal{E}^2 x_4^2
+ 
E^2(\mathcal{E})
\quad
\Rightarrow
\quad
G
= \frac{1}{2 \mathcal{H} + E^2(\mathcal{E})}
,\end{equation}
where 
$\mathcal{H}$ 
is the harmonic oscillator Hamiltonian of the auxiliary quantum mechanics. 
The two-point function can then be computed as a function of 
$\tau$
using a method well known to those who know it well: 
Schwinger's proper-time trick~\cite{Schwinger:1951nm,Tiburzi:2008ma}. 
The result is, of course, not a simple exponential in time; 
but, 
the predicted form of the correlation function matches well against lattice data~\cite{Detmold:2009dx},
and can be used to extract the polarizability.

When one considers spin-$\frac{1}{2}$ hadrons, 
an additional ingredient appears, namely the magnetic moment interaction. 
This is relevant even for an external electric field because the interaction has the form
$\sigma_{\mu \nu} F_{\mu \nu}$
and does not disappear from the effective action when the velocity is projected to zero. 
An unpolarized neutron correlation function, 
$\texttt{Tr} [ G_{\mathcal{E}}(\tau)]$ 
is described at long times by an exponential 
fall-off, 
but governed by
$E(\mathcal{E}) = M_n + \frac{1}{2} \mathcal{E}^2 [ \alpha^n_E - \frac{\mu_n^2}{4 M_n^3}]$.  
One must separate out the magnetic moment contribution to determine the electric polarizability. 
This can be achieved with so-called boost projection~\cite{Detmold:2010ts}, 
which utilizes the information contained in the various spin-components of the correlation function. 
For a magnetic field, 
one has 
$\sigma_{\mu \nu} F_{\mu \nu} = \vec{\sigma} \cdot \vec{B}$
and one uses spin projection to separate out the two energy levels. 
On the other hand, 
for an electric field, 
$\sigma_{\mu \nu} F_{\mu \nu} = \vec{K} \cdot \vec{E}$, 
where 
$\vec{K}$ 
are the boost generator matrices. 
Tracing with boost projectors gives a way to separate out the magnetic moment
\begin{equation}
\texttt{Tr} [ ( 1 \pm K_3 ) G_{\mathcal{E}}(\tau)] 
= 
Z_{\mathcal{E}} 
\left( 1 \pm \frac{\mu_n \mathcal{E}}{2 M_n} \right)
e^{ - E(\mathcal{E} ) \tau}
,\end{equation}
but it is done from the differing amplitudes. 
The boost-projection method has also been extended to proton correlation functions~\cite{Detmold:2010ts}.





\section{Outlook}

The background field method is a practical current-day technique to compute the polarizabilities of hadrons. 
Our study was not limited to spin-less neutral hadrons. 
On the contrary, 
we developed new techniques to handle charged hadrons, and spin-$\frac{1}{2}$ hadrons. 
Although we pursued computations in background electric fields, 
such methods are easily generalized to magnetic fields. 
Lattice sizes are nearly large enough to support perturbatively small values of quantized magnetic fields for the study of 
magnetic moments and magnetic polarizabilities. 
There are a number of areas for improvement in our calculations. 
We intend to study the pion mass dependence of the extracted polarizabilities to make contact with chiral perturbation theory. 
This can be done even with electrically neutral sea quarks, 
as partially quenched chiral perturbation theory has been employed to determine the sea quark charge dependence of 
nucleon and pion polarizabilities~\cite{Detmold:2006vu,Hu:2007ts}.  
Another notable area of improvement and concern is the effect of finite volume on the extraction of polarizabilities. 
On a periodic lattice, 
virtual pions that propagate around the world $n$-times lead to finite volume artifacts that scale as
$e^{ - n m_\pi L}$. 
For our background field simulations, 
contributions from virtual charged pions wrapping around the world
$n$-times (plus $n$-anti-times) are accompanied by Wilson lines leading 
to modified finite volume effects of the form
$e^{- n m_\pi L} \cos ( n \Phi )$~\cite{Tiburzi:2008pa,Detmold:2009fr}.  
Here $\Phi$ is the holonomy of the gauge field, 
which, 
for us, 
reads
$\Phi = \int_0^L A_3 dz = - \mathcal{E} L x_4$. 
The presence of such time-dependent interactions complicates the extraction of the desired infinite volume physics from two-point functions. 
We are investigating means to handle such volume corrections using single-particle effective actions in time-dependent perturbation theory.\footnote{
An alternate way to  handle such effects is to perturbatively expand the background field method from the start. 
This gives one direct control over Wilson loop contributions which appear only as $\propto \Phi^2$, 
and can be separated out from the polarizability 
signal by varying the reference time \cite{Engelhardt:2007ub}.  
}
Nonetheless, 
the lattice can make a fundamental contribution to hadron physics through the computation of  polarizabilities. 
Such computations will be timely because improved experimental measurements are anticipated. 
The results, 
moreover, 
will allow us to test our understanding of low-energy QCD from first principles.


\begin{theacknowledgments}
Work supported in part by Jefferson Science Associates, LLC under U.S.~Dept.~of Energy contract No.~DE-AC05-06OR-23177 (W.D.), 
and the U.S.~Dept.~of Energy, 
under Grant Nos.~DE-SC000-1784 (W.D.), 
DE-FG02-94ER40818 (B.C.T.), 
and 
DE-AC02-05CH11231 (A.W.-L.).  
\end{theacknowledgments}



\bibliographystyle{aipproc}   


\begin{thebibliography}{99}

\bibitem{Chodos:1974je}
  A.~Chodos, R.~L.~Jaffe, K.~Johnson, C.~B.~Thorn, V.~F.~Weisskopf,
  Phys.\ Rev.\  {\bf D9}, 3471-3495 (1974).

\bibitem{Schafer:1984hw}
  A.~Sch\"afer, B.~M\"uller, D.~Vasak, W.~Greiner,
  Phys.\ Lett.\  {\bf B143}, 323-325 (1984).

\bibitem{Gasser:1983yg}
  J.~Gasser, H.~Leutwyler,
  Annals Phys.\  {\bf 158}, 142 (1984).

\bibitem{Holstein:1990qy}
  B.~R.~Holstein,
  Comments Nucl.\ Part.\ Phys.\  {\bf A19}, 221-238 (1990).

\bibitem{Bernard:1991rq}
  V.~Bernard, N.~Kaiser, U.~G.~Meissner,
  Phys.\ Rev.\ Lett.\  {\bf 67}, 1515-1518 (1991).

\bibitem{Ahrens:2004mg}
  J.~Ahrens, V.~M.~Alexeev, J.~R.~M.~Annand, H.~J.~Arends, R.~Beck, G.~Caselotti, S.~N.~Cherepnya, D.~Drechsel {\it et al.},
  Eur.\ Phys.\ J.\  {\bf A23}, 113-127 (2005).
  [nucl-ex/0407011].

\bibitem{Bernard:1982yu}
  C.~W.~Bernard, T.~Draper, K.~Olynyk, M.~Rushton,
  Phys.\ Rev.\ Lett.\  {\bf 49}, 1076 (1982).

\bibitem{Martinelli:1982cb}
  G.~Martinelli, G.~Parisi, R.~Petronzio, F.~Rapuano,
  Phys.\ Lett.\  {\bf B116}, 434 (1982).
  
\bibitem{Fiebig:1988en}
  H.~R.~Fiebig, W.~Wilcox, R.~M.~Woloshyn,
  Nucl.\ Phys.\  {\bf B324}, 47 (1989).
  
\bibitem{'tHooft:1979uj}
  G.~'t Hooft,
  Nucl.\ Phys.\  {\bf B153}, 141 (1979).  
  
\bibitem{Detmold:2008xk}
  W.~Detmold, B.~C.~Tiburzi, A.~Walker-Loud,
  PoS {\bf LATTICE2008}, 147 (2008).
  [arXiv:0809.0721 [hep-lat]].
  
\bibitem{Detmold:2009dx}
  W.~Detmold, B.~C.~Tiburzi, A.~Walker-Loud,
  Phys.\ Rev.\  {\bf D79}, 094505 (2009).
  [arXiv:0904.1586 [hep-lat]].
  
  
\bibitem{Detmold:2006vu}
  W.~Detmold, B.~C.~Tiburzi, A.~Walker-Loud,
  Phys.\ Rev.\  {\bf D73}, 114505 (2006).
  [hep-lat/0603026].
  
 
\bibitem{Schwinger:1951nm}
  J.~S.~Schwinger,
  Phys.\ Rev.\  {\bf 82}, 664-679 (1951).

 
\bibitem{Tiburzi:2008ma}
  B.~C.~Tiburzi,
  Nucl.\ Phys.\  {\bf A814}, 74-108 (2008).
  [arXiv:0808.3965 [hep-ph]].

  
\bibitem{Detmold:2010ts}
  W.~Detmold, B.~C.~Tiburzi, A.~Walker-Loud,
  Phys.\ Rev.\  {\bf D81}, 054502 (2010).
  [arXiv:1001.1131 [hep-lat]].


\bibitem{Hu:2007ts}
  J.~Hu, F.~-J.~Jiang, B.~C.~Tiburzi,
  Phys.\ Rev.\  {\bf D77}, 014502 (2008).
  [arXiv:0709.1955 [hep-lat]].


\bibitem{Tiburzi:2008pa}
  B.~C.~Tiburzi,
  Phys.\ Lett.\  {\bf B674}, 336-343 (2009).
  [arXiv:0809.1886 [hep-lat]].

\bibitem{Detmold:2009fr}
  W.~Detmold, B.~C.~Tiburzi, A.~Walker-Loud,
  [arXiv:0908.3626 [hep-lat]].


\bibitem{Engelhardt:2007ub}
  M.~Engelhardt [ LHPC Collaboration ],
  Phys.\ Rev.\  {\bf D76}, 114502 (2007).
  [arXiv:0706.3919 [hep-lat]].




\end{thebibliography}



\end{document}